# The necessities for building a model to evaluate Business Intelligence projects- Literature Review

Vahid Farrokhi[1] and László Pokorádi[2]

[1, 2]Doctoral School of Informatics, University of Debrecen, Debrecen, Hungary

[1]vahid.farrokhi@inf.unideb.hu and [2]pokoradi@eng.unideb.hu

*Abstract*

*In recent years Business Intelligence (BI) systems have consistently been rated as one of the highest priorities of Information Systems (IS) and business leaders. BI allows firms to apply information for supporting their processes and decisions by combining its capabilities in both of organizational and technical issues. Many of companies are being spent a significant portion of its IT budgets on business intelligence and related technology. Evaluation of BI readiness is vital because it serves two important goals. First, it shows gaps areas where company is not ready to proceed with its BI efforts. By identifying BI readiness gaps, we can avoid wasting time and resources. Second, the evaluation guides us what we need to close the gaps and implement BI with a high probability of success. This paper proposes to present an overview of BI and necessities for evaluation of readiness.*

## 1. Introduction

Many firms have realized that the only way to continually compete and profit in the global marketplace of today is to utilize the power of information [1-3]. In today's highly competitive world, the quality and timeliness of business information for an organization is not just a choice between profit and loss; it may be a question of survival or bankruptcy [4] . The business needs to know what is happening right now, faster, in order to determine and influence what should happen next time [5]. Companies spend billions of dollars annually on implementation and maintenance of IS [6]. Estimates are that IS expenses constitute the largest portion of organizational expenditures [7, 8]. Given the size of these expenditures, companies expect to gain benefits commensurate with the money being spent. Unfortunately recent figures estimated that nearly half of IS project did not result in the anticipated benefits [8]. So it is important to know how companies can get a benefit and suitable return on their investments.

Previous information systems like maintaining accounting ledgers or processing financial transactions were applied to automate manual processes. The benefits from these types of systems resulted from increases in efficiency or effectiveness of the underlying processes resulting in measurable cost saving or revenue increases [9]. Traditional enterprises may normally face issues such as the overflow of data, the lack of information, the lack of knowledge and insufficiency of reports [10]. Top managers used to make and take decisions based on their experiences which these lead to more risk of decision failure and reducing the value of the decision. As worldwide competition is maturing, past decision-making modes can no longer satisfy the requirements of enterprises for decision efficiency and benefits; enterprises must make good use of electronic tools to quickly extract useful information from huge volume of data by providing the skills of fast decision-making [11]. Socio-economic reality of contemporary organizations has made organizations face some necessity to look for instruments that would facilitate effective acquiring, processing and analyzing vast amounts of data that come from different and dispersed sources and that would serve as some basis for discovering new knowledge [12]. Recent years, there are many software packages which can provide a set of complete solutions for the operation and management processes of organizations. Nowadays, the individual-system approach applied to decision-support such as Decision Support Systems (DSS) has been substituted by a new environmental approach [13]. With the potential to gain competitive advantage when making important decisions, it is vital to integrate decision support into the environment of their enterprise and work systems. Business Intelligence can be embedded in these enterprise systems to obtain this competitive advantage [14, 15]. BI systems provide benefits by supporting analytical processes that provide recommendations for changing products or processes in ways that improve their competitiveness or operational efficiency [16]. And practitioners design and implement Business Intelligence as umbrella concept create a decision-support environment for management in enterprise systems [17].

However, the effects of the implementation of electronization tools vary that the probability of failure is higher than that of the success [18]. Therefore, the ability to implement BI project and support it, depends on readiness of companies.

Farrokhi and Suhaimi in their article addressed about importance of information and its flow in companies and developed a model based on Business Process Modeling (BPM) [19]. Business intelligence technology gives this ability to the managers and experts of these companies. But nowadays BI systems include one of the largest and fastest growing areas of IT expenditure in companies and if the BI project fail, they will lose a lot of money. For reducing costs through BI implementation and preventing from fail of BI project, we need to evaluate readiness of these companies from two aspects: Organizational and Technical.

This paper presents an overview of BI and necessities for building a model which enables us to assay and evaluate readiness of companies from technical and organizational aspects when they want to implement BI project. The outline of the paper is as follows: Section 2 shows an overview of BI and its definitions. Section 3 presents the necessities for building an evaluation model of readiness. Finally, section 4 presents the conclusions and prospective.

## 2. Business Intelligence

Business Intelligence or BI is a grand, umbrella term, introduced by Howard Drenser of the Gartner Group, in 1989, to describe a set of concepts and methods to improve business decision making by using fact-based, computerized support systems [20]. The first scientific definition by Ghoshal and kim [21] referred to BI as a management philosophy and tool that helps organizations to manage and refine business information for the purpose of making effective decisions. The goal of BI systems [5] is to capture (data, information, knowledge) and respond to business events and needs better, more informed, and faster, as decisions. BI was considered to be an instrument of analysis, providing automated decision making about business conditions, sales, customer demand, product preference and so on [22]. The Data Warehousing Institute, a provider of education and training in data warehouse and BI industry defines business intelligence as: The processes, technologies, and tools needed to turn data into information, information into knowledge, and knowledge into plans that drive profitable business action. Business intelligence encompasses data warehousing, business analytic tools, and content/knowledge management.[1] Business intelligence has been defined as "business information and business analyses within the context of key business processes that lead to decisions and actions and that result in improved business performance" [23]. Another definitions is "a set of processes and technologies that transform raw, meaningless data into useful and actionable information" [24]. It utilizes a substantial amount of collected data during the daily operational processes, and transforms the data into information and knowledge to avoid

---

[1] The Data Warehouse Institute Faculty Newsletter, Fall 2002.

the supposition and ignorance of the enterprises [25]. Golfarelli at al. argue that BI is the process that transforms data into information and then into knowledge [26]. It is the process of gathering high-quality and meaningful information about the subject matter being researched that will help the individual(s) to analyze the information, draw conclusions or make assumptions [27]. Stackowiak et al. opine that BI is the process of taking large amounts of data, analyzing that data, and presenting a high-level set of reports that condense the essence of that data into the basis of business actions, enabling management to make fundamental daily business decisions [28]. Zeng et al. have put forth that BI is "The process of collection, treatment and diffusion of information that has an objective, the reduction of uncertainty in the making of all strategic decisions [29]. Ranjan [30] considers BI as the conscious methodical transformation of data from any and all data sources into new forms to provide information that is business-driven and results-oriented. Eckerson [31] understood that BI must be able to provide the following tools: production reporting, end-user query and reporting, OnLine Analytical Processing (OLAP), dashboard/screen tools, data mining tools, and planning and modeling tools. It uses huge-database (data-warehouse) analysis, and mathematical, statistical and artificial intelligence, as well as data mining and OLAP [32]. BI includes a set of concepts, methods and processes to improve business decisions, using information from multiple sources and applying past experience to develop an exact understanding of business dynamics [33]. It has emerged as a concept for analyzing collected data with the purpose to help decision making units get a better comprehensive knowledge of an organization's operations, and thereby make better business decisions [4]. A BI system is a data-driven DSS that primarily supports the querying of a historical database and the production of periodic summary reports [34]. It can be presented as an architecture, tool, technology or system that gathers and stores data, analyzes it using analytical tools, facilities reporting, querying and delivers information and/or knowledge that ultimately allows organizations to improve decision making [35-42]

Lönnqvist and Pirttimäki [43] stated that term, BI, can be used when referring to the following concepts:
1. Related information and knowledge of an organization, which describe the business environment, the organization itself, the conditions of the market, customers and competitors and economic issues;
2. Systemic and systematic processes by which organizations obtain, analyse and distribute the information for making decisions about business operations.

BI allows firms to apply information for supporting their processes and decisions by combining its capabilities in both of organizational and technical issues. Put another way, "business intelligence allows people at all levels of an organization to access, interact with, and analyze data to manage the business, improve performance, discover opportunities, and operate efficiently" [44]. Problems and a huge amount of data of enterprises are input into data mining systems for data analysis so that decision makers can obtain useful information promptly for making correct judgment; that is, in regard to enterprise operating contents, abilities of fast understanding and deducing are provided, and thus enhancing the quality of decision-making and improving performance and expediting processing speed [45]. From a technical perspective, BI systems offer an integrated set of tools, technologies and software products that are used to collect heterogenic data from dispersed sources in order to integrate and analyse data to make it commonly available [12].

In some research, BI is concerned with the integration and consolidation of raw data into key performance indicators (KPIs). KPIs represent an essential basis for business decisions in the context of process execution. Therefore, operational processes provide the context for data analysis, information interpretation, and the appropriate action to be taken [46]. Figure 1 depicts this concept.

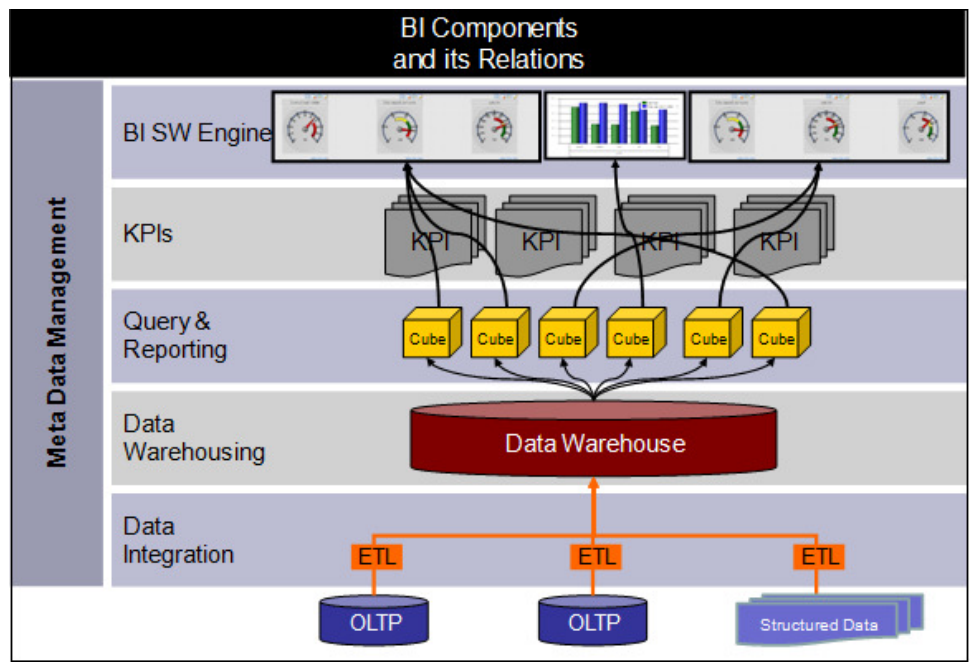

Figure 1: KPIs and BI Components (Source: [47])

Therefore, BI covers a wide range of tools and broad scope, and among the commonly mentioned important applications are data warehouse, data mining, OLAP, DSS, Balance Scorecard (BSC), etc [10]. However, in the overall view, there are two important issues. First, the core of BI is the gathering, analysis and distribution of information. Second, the objective of BI is to support the strategic decision-making process [22]. By strategic decisions, we mean decisions related to implementation and evaluation of organizational vision, mission, goals and objectives with medium to long-term impact on the organization, as opposed to operational decisions, which are day-to-day in nature and more related to execution [48].

## 3. Necessities for evaluation of readiness

In recent years Business Intelligence systems have consistently been rated as one of the highest priorities of IS and business leaders [24, 49, 50]. Winning companies, such as Continental Airlines, have seen investments in BI generate increases in revenue and produce cost savings equivalent to a 1,000% return on investment (ROI) [51]. Many of companies are being spent a significant portion of its IT budgets on business intelligence and related technology. Estimates of the amount spent on BI in 2006 range from $14 to $20 Billion, with growth estimates of from 10% to 11% per year for the foreseeable future [44, 52]. A Gartner Executive Program survey, as shown in figure 2, conducted in 2008 across 1,500 organizations in Western Europe found that BI is the top technology priority for CIOs.

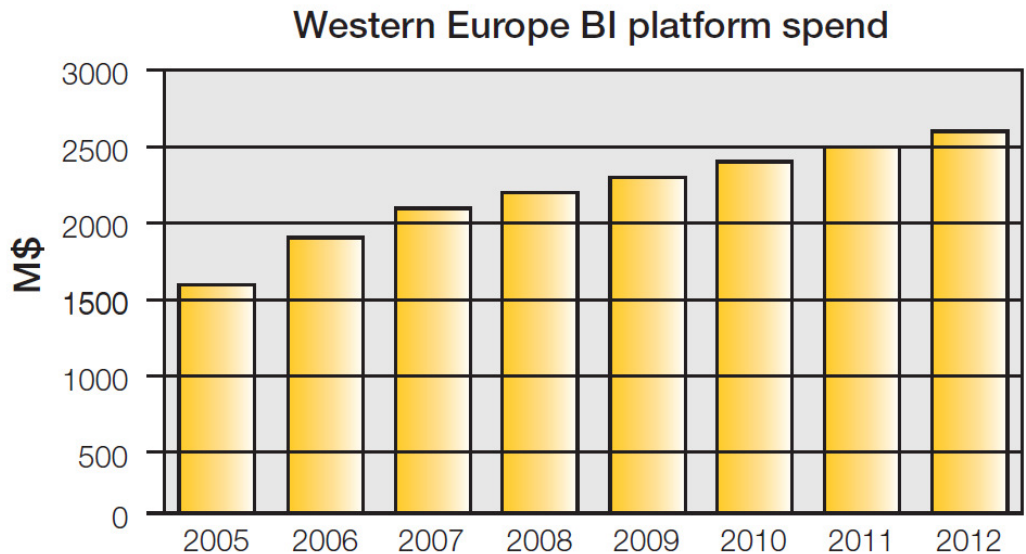

Figure 2: BI Spend Prediction (Source: [53])

In spite of these investments only 24% of BI implementations were identified as being very successful in a recent survey of companies using BI systems [44]. Losing companies have spent more resources than their competitors with a smaller ROI, all while watching their market share and customer base continuously shrink [54]. The complexity of business intelligence – data warehouse systems is very high so it is better to consider from the beginning various foreseen aspects that could impact the overall cost and increase the initial investment of the project but even with a good analysis there are still remaining a large numbers of variables to be considered [55]. The companies which are investing heavily in BI must expect to achieve benefits from their investments. How can some organizations achieve to these benefits while others don't? What are differences between those companies which gain benefits from BI implementation with the companies lost their money? Unfortunately, while much has been written about how to effectively implement and use business intelligence technology [23, 44, 56, 57], research on BI and specifically detailing how an organization can achieve benefits from BI is sparse [58].

### 3.1. BI Success

The stakes are high for organizations to develop successful BI implementations [59]. When we want to research how BI can be considered successful we have to be able to define what we mean by success. As we know, BI is a class of information system and it is better that we begin to clarify how success is measured for IS in general. Many IS researchers have tried to evaluate success [60-65]. Early work focused on multiple criteria including "profitability, application to major problems of the organization, quality of decisions or performance, user satisfaction and wide-spread use" [61]. The appropriate success measure depended upon the perspective of those evaluating success or the nature of the problem being addressed [66].

While multiple criteria measures are useful in IS success but many of those criteria are difficult to measure. As a result, much of the work on IS success has focused on system use as a proxy for success [67-69]. In other words, the authors advised that capability of system usage is an important clue for its success. Usage of an information system means that the system can be accepted by users, and users' work-related needs can be met and the objective at the initial implementation can be achieved. Still it was recognized that "a better measure of IS success would probably be some weighted average of the criteria" [61]. So the advantages of an information system differ and it depends on the type of system being implemented and the stakeholder of it. This guides us that success measures for the research is necessary to be based on BI specific characteristics. BI systems are implemented to provide analytical capability to offer recommendations to improve operational or strategic processes or product characteristics [23, 44]. Value of BI for business is predominantly expressed in the fact that such systems cast some light on information that may serve as the basis for carrying out fundamental changes in a particular enterprise, i.e. establishing new co-operation, acquiring new customers, creating new markets, offering products to customers [70-72]. This means that using a BI system is not enough to say it is successful but using recommendations and advices is the key factor. Thus we should consider the achievement of organizational benefits to be appropriate measure of BI success.

### 3.2. BI Readiness

BI readiness means that the essential prerequisites for BI success are in place. BI readiness assessments are used at the front end of BI projects to determine the degree to which a given company is prepared to make the changes that are necessary to capture the full business value of BI [23]. The BI Readiness Assessment is a series of tasks that analyzes several key areas across an organization to evaluate how prepared an organization is to begin short term tactical deployment of Business Intelligence solutions and mature it practice over the long term [73]. Evaluation of BI readiness is vital because it serves two important goals. First, it shows gaps areas where company is not ready to proceed with its BI efforts. By identifying BI readiness gaps, we can avoid wasting time and resources. Second, the evaluation guides us what we need to close the gaps and implement BI with a high probability of success.

### 3.3. Necessities for building a model

The bottom line in any evaluation program is the finding of problems and the demonstration that the system under evaluation satisfies its requirements. It is unfortunate that, in many cases, the evaluating program is actually aimed at showing that the BI system, as implemented, runs as it is requested by the users. That is, the evaluations are aimed at showing that the BI project does not fail, rather than that it fulfills its requirements.

There are a few books that discuss exactly on BI readiness. Williams and Williams (2007) identified seven factors defining "business intelligence readiness" as being:

  i. Strategic Alignment;
 ii. Continuous Process Improvement Culture;
iii. Culture Around Use of Information and Analytics;
 iv. BI Portfolio Management;
  v. Decision Process Engineering Culture;
 vi. BI & DW Technical Readiness;
vii. Business/IT Partnership [23].

The authors (S. Williams, and Williams, N.) suggested that only when an organization can gain the benefits of BI, if it has this readiness. Davenport and Harris in their book "Competing on Analytics," [56] focused on the impact of BI systems on organizations. They identified something that called an analytical capability, which was their conception of the ability of an organization to use BI and as consisting of organizational acumen and technology factors [56]. They suggest that an organization need to have capability in both organizational and technology factors. But they provide a high level view of these factors without discuss in detail.

Jourdan *et al.* have collect, synthesizes, and analyzes 167 articles on a variety of topics closely related to business intelligence published from 1997 to 2006 in ten leading Information Systems journals [59]. Based on their research, there are only 35 articles in BI implementation category which covers issues in a variety of BI contexts including data warehousing, data mining, Customer Relationship Management (CRM), Enterprise Resource Planning (ERP), Knowledge Management Systems (KMS), and eBusiness projects.

Research in information systems is generally focused on either developing theories that explain related phenomena or on verifying existing theories [74]. Analysis of the research strategies (in BI Research) over the ten year period from 1997 to 2006 illustrates that Formal Theory/Literature Review, Field Study-Primary Data, Field Study-Secondary Data, and Sample Survey are represented in almost every year of the time frame [59]. These four strategies are exploratory in nature and indicate the beginnings of a body of research [75]. BI research covers diverse subjects ranging from practical applications of neural networks [76], to end-user satisfaction [77], to the use of clustering as a business strategy to gain a competitive advantage [78].

Based on the journals and the books mentioned above and previous sections, there is not any research of evaluation of BI readiness in companies. So we need to:
  i.  investigate and determine BI readiness factors and their associated contextual elements that influence implementation of BI systems in companies
  ii. developing a model for evaluation of BI readiness in companies

## 4. Conclusions and Prospective

In this paper an attempt has been made to depict an overview of BI and the necessities for building a model to evaluate readiness of companies in implementing BI project. It was shown that in today's highly competitive world, using BI is vital and no business organization can deny the benefit of BI. BI technologies are applied by profit and non-profit firms and business users became increasingly proactive. Successful BI project is an important issue for both researchers and practitioners; however, not many studies have done on BI readiness. Although some guidelines for implementation exist, few have been subjected to model building. During prospective scientific research related to this study, the authors will work out models to evaluate readiness of companies in implementing BI projects.